\documentclass[12pt,italian,english]{article}
\usepackage[T1]{fontenc}
\usepackage[latin9]{inputenc}
\usepackage{amssymb}
\usepackage{graphicx}

\makeatletter
\newcommand{\lyxaddress}[1]{
\par {\raggedright #1
\vspace{1.4em}
\noindent\par}
}

\makeatother

\usepackage{babel}
\begin{document}

\title{\textbf{Differential geometry and scalar gravitational waves }}

\author{\textbf{Christian Corda}}

\maketitle

\lyxaddress{\begin{center}
Institute for Theoretical Physics and Mathematics Einstein-Galilei,
Via Santa Gonda, 14 - 59100 Prato, Italy
\par\end{center}}
\begin{abstract}
Following some strong argumentations of differential geometry in the
Landau's book, some corrections about errors in the old literature
on scalar gravitational waves (SGWs) are given and discussed. 

In the analysis of the response of interferometers the computation
is first performed in the low frequencies approximation, then the
analysis is applied to all SGWs in the full frequency and angular
dependences. 

The presented results are in agreement with the more recent literature
on SGWs. 
\end{abstract}

\lyxaddress{PACS numbers: 04.80.Nn, 04.30.Nk, 04.50.+h}

\bigskip{}

\bigskip{}

\bigskip{}

\bigskip{}

\section{Introduction}

Second generation interferometric GW detectors, such as Advanced LIGO,
are expected to begin operation by 2015 \cite{key-1}. The realization
of a GW astronomy, by giving a significant amount of new information,
will be a cornerstone for a better understanding of gravitational
physics. In fact, the discovery of GW emission by the compact binary
system PSR1913+16, composed by two Neutron Stars \cite{key-2}, has
been, for physicists working in this field, the ultimate thrust allowing
to reach the extremely sophisticated technology needed for investigating
in this field of research. GW astronomy plans to reach sensitivities
that will permit to test General Relativity in the dynamical, strong
field regime and investigate departures from its predictions, becoming,
alternatively, a strong endorsement for Extended Theories like $f(R)$
Theories or Scalar Tensor Gravity \cite{key-3}.

While the response of interferometers to standard tensor GWs in General
Relativity has been calculated in lots of works (see for example \cite{key-3}-\cite{key-9}),
the interaction between interferometers and SGWs arising from Scalar
Tensor Gravity is a more recent field of interest and it has not an
analogous number of works in the literature. The first work of \cite{key-10}
was improved by the work of the authors of \cite{key-11}. In \cite{key-10}
the authors did not realize that in their gauge the beam splitter
is not left at the origin by the passage of the SGW and furthermore
computed a coordinate-time interval than a proper-time interval, reaching
the incorrect conclusion that the SGW has longitudinal effect, and
does not have transverse one. In \cite{key-11} the transverse effect
of SGWs was shown in the gauge in \cite{key-10}. After this, in \cite{key-12},
the analysis of \cite{key-11} was generalized with the computation
of the frequency-dependent angular pattern of interferometers in the
gauge of \cite{key-10}, while in \cite{key-11} the angular pattern
was only computed in the low frequencies approximation (wavelength
much larger than the linear dimensions of the interferometer, under
this assumption the amplitude of the SGW, $\Phi$, can be considered
''frozen'' at a value $\Phi_{0}$ ).

In this paper the gauge in \cite{key-10} is re-analysed. Following
some strong argumentations on differential geometry in the Landau's
book \cite{key-13}, we show that in \cite{key-10} there were errors
in the geodesic equations of motion too. These errors reflected also
in \cite{key-11} and \cite{key-12} where incorrect equation of motion
taken from \cite{key-10} were used. Thus, we correct erroneous results
which were in three papers published in Physical Review D \cite{key-10,key-11,key-12}.

In the analysis of the response of interferometers the computation
is first made in the low frequencies approximation like in \cite{key-11},
then the calculation is generalized to all SGWs. 

At the end of this paper, the correct detector pattern of interferometers
in the gauge in \cite{key-10} is computed. The presented results
are in full agreement with the recent ones in \cite{key-3,key-14,key-15,key-16}.

\section{A particular gauge for scalar gravitational waves}

We consider a gauge which was proposed in the first time in \cite{key-10}.
In this gauge a purely plane SGW is travelling in the $z+$ direction
(progressive wave $\Phi\equiv\Phi(t-z)$) and acting on an interferometer
whose arms are aligned along the $x$ and $z$ axes \cite{key-3,key-10,key-11,key-14}).
In this gauge it is \cite{key-10,key-11,key-14}

\begin{equation}
e_{\mu\nu}^{(s)}=\eta_{\mu\nu}.\label{eq: gauge di Shibata}
\end{equation}

Thus, the line element results the conformally flat one (we work with
$c=1$ and $\hbar=1$ in this paper)

\begin{equation}
ds^{2}=[1+\Phi(t-z)](-dt^{2}+dz^{2}+dx^{2}+dy^{2}).\label{eq: metrica puramente scalare di Shibata}
\end{equation}

Eq. (\ref{eq: gauge di Shibata}) can be rewritten as

\begin{equation}
(\frac{dt}{d\tau})^{2}-(\frac{dx}{d\tau})^{2}-(\frac{dy}{d\tau})^{2}-(\frac{dz}{d\tau})^{2}=\frac{1}{(1+\Phi)},\label{eq: Sh2}
\end{equation}

where $\tau$ is the proper time of the test masses.

From eqs. (\ref{eq: metrica puramente scalare di Shibata}) and (\ref{eq: Sh2})
the authors of \cite{key-10} obtained the geodesic equations of motion
for test masses (i.e. the beam-splitter and the mirrors of the interferometer),
see eq. 3.21, 3.22 and 3.23 of \cite{key-10},

\begin{equation}
\begin{array}{ccc}
\frac{d}{d\tau}[(1+\Phi)\frac{dx}{d\tau}] & = & 0\\
\\
\frac{d}{d\tau}[(1+\Phi)\frac{dy}{d\tau}] & = & 0\\
\\
\frac{d}{d\tau}[(1+\Phi)\frac{dt}{d\tau}] & = & \frac{1}{2}\frac{\partial_{t}(1+\Phi)}{(1+\Phi)}\\
\\
\frac{d}{d\tau}[(1+\Phi)\frac{dz}{d\tau}] & = & -\frac{1}{2}\frac{\partial_{z}(1+\Phi)}{(1+\Phi)}.
\end{array}\label{eq: geodetiche Shibata}
\end{equation}

By using some strong argumentations on differential geometry in the
Landau's book \cite{key-13}, below we will show that eqs. (\ref{eq: geodetiche Shibata})
are not correct.

Other incorrect geodesic equations of motion were used in \cite{key-12},
see eqs. 4.2, 4.3, 4.4 and 4.5, in this case for a wave travelling
in the $z-$direction (regressive wave). Therefore, the results of
\cite{key-12} and in particular the frequency-dependent angular pattern
of eq. (5.25) are not correct.

To derive the correct geodesic equation of motion for a progressive
wave, eq. (87,3) in \cite{key-13}, which is
\begin{equation}
\frac{d^{2}x^{i}}{d\tau^{2}}+\Gamma_{kl}^{i}\frac{dx^{k}}{d\tau}\frac{dx^{l}}{d\tau}=0,\label{eq: LL}
\end{equation}

can be used.

In this way, using equation (\ref{eq: Sh2}), one gets

\begin{equation}
\begin{array}{ccc}
\frac{d^{2}x}{d\tau^{2}} & = & 0\\
\\
\frac{d^{2}y}{d\tau^{2}} & = & 0\\
\\
\frac{d^{2}t}{d\tau^{2}} & = & \frac{1}{2}\frac{\partial_{t}(1+\Phi)}{(1+\Phi)^{2}}\\
\\
\frac{d^{2}z}{d\tau^{2}} & = & -\frac{1}{2}\frac{\partial_{z}(1+\Phi)}{(1+\Phi)^{2}}.
\end{array}\label{eq: geodetiche Corda}
\end{equation}

Now, following \cite{key-13}, we show the difference between the
correct eqs. (\ref{eq: geodetiche Corda}) and the incorrect ones
(\ref{eq: geodetiche Shibata}). Let us review the important demonstration
in Chapter 10, Paragraph 86 of \cite{key-13}, which implies that
the covariant derivative of the metric tensor is equal to zero, i.e.
the metric tensor works like a constant in the covariant derivative.
Calling $DA_{i}$ the covariant derivative of an arbitrary vector
$A_{i}$, it is \cite{key-13} 
\begin{equation}
Du_{i}=g_{ik}Du^{k},\label{eq: uno}
\end{equation}
but it is also \cite{key-13}

\begin{equation}
Du_{i}=D(g_{ik}u^{k})=(Dg_{ik})u^{k}+g_{ik}Du^{k}.\label{eq: due}
\end{equation}
 Eqs. (\ref{eq: uno}) and (\ref{eq: due}) imply 

\begin{equation}
Dg_{ik}=0.\label{eq: differenziale}
\end{equation}

On the other hand, eqs. (\ref{eq: LL}) arise from eq. (87,2) in \cite{key-13},
which is

\begin{equation}
Du^{i}=0.\label{eq: geodetiche iniziali}
\end{equation}

By combining eqs. (\ref{eq: geodetiche iniziali}) with eqs. (\ref{eq: differenziale})
one realizes immediately that the components of the metric tensor,
which are all equal to $(1+\Phi)$ in the line element (\ref{eq: metrica puramente scalare di Shibata}),
\textbf{have to be put outside the derivatives in eqs. (\ref{eq: geodetiche Shibata})}.
In other words, eqs. (\ref{eq: geodetiche Shibata}) have to be rewritten
as 

\begin{equation}
\begin{array}{ccc}
(1+\Phi)\frac{d^{2}x}{d\tau^{2}} & = & 0\\
\\
(1+\Phi)\frac{d^{2}y}{d\tau^{2}} & = & 0\\
\\
(1+\Phi)\frac{d^{2}t}{d\tau^{2}} & = & \frac{1}{2}\frac{\partial_{t}(1+\Phi)}{(1+\Phi)}\\
\\
(1+\Phi)\frac{d^{2}z}{d\tau^{2}} & = & -\frac{1}{2}\frac{\partial_{z}(1+\Phi)}{(1+\Phi)}.
\end{array}\label{eq: shibata corrette}
\end{equation}

that become

\begin{equation}
\begin{array}{ccc}
\frac{d^{2}x}{d\tau^{2}} & = & 0\\
\\
\frac{d^{2}y}{d\tau^{2}} & = & 0\\
\\
\frac{d^{2}t}{d\tau^{2}} & = & \frac{1}{2}\frac{\partial_{t}(1+\Phi)}{(1+\Phi)^{2}}\\
\\
\frac{d^{2}z}{d\tau^{2}} & = & -\frac{1}{2}\frac{\partial_{z}(1+\Phi)}{(1+\Phi)^{2}}.
\end{array}\label{eq: confronto 2}
\end{equation}

These last equations are exactly eqs. (\ref{eq: geodetiche Corda}).

In \cite{key-10} the authors did not take in due account the important
issue of differential geometry that every component of the metric
tensor works like a constant in the covariant derivative. Instead,
previous analysis will be fundamental for the conservation of some
quantities in the next analysis.

The first and the second of eqs. (\ref{eq: geodetiche Corda}) can
be immediately integrated obtaining

\begin{equation}
\frac{dx}{d\tau}=C_{1}=const.\label{eq: integrazione x}
\end{equation}

\begin{equation}
\frac{dy}{d\tau}=C_{2}=const.\label{eq: integrazione y}
\end{equation}

In this way eq. (\ref{eq: Sh2}) becomes
\begin{equation}
(\frac{dt}{d\tau})^{2}-(\frac{dz}{d\tau})^{2}=\frac{1}{(1+\Phi)}.\label{eq: Ch3}
\end{equation}

Assuming that test masses are at rest initially it is $C_{1}=C_{2}=0$.
Thus, even if the SGW arrives at test masses, there is not motion
of test masses within the $x-y$ plane in this gauge. 

Now, we show that, in presence of a SGW, there is motion of test masses
in the $z$ direction which is the direction of the propagating wave.
An analysis of eqs. (\ref{eq: geodetiche Corda}) shows that, to simplify
equations, the retarded and advanced time coordinates ($u,v$) can
be introduced, exactly like in \cite{key-10}

\begin{equation}
\begin{array}{c}
u=t-z\\
\\
v=t+z.
\end{array}\label{eq: ret-adv}
\end{equation}

From the third and the fourth of eqs. (\ref{eq: geodetiche Corda})
we get

\begin{equation}
\frac{d}{d\tau}\frac{du}{d\tau}=\frac{\partial_{v}[1+\Phi(u)]}{(1+\Phi(u))^{2}}=0.\label{eq: t-z t+z}
\end{equation}

Equation (\ref{eq: t-z t+z}) represents the fundamental difference
with the work of \cite{key-10}. The authors of \cite{key-10} found
the equation (see eq. (3.27) of \cite{key-10})
\begin{equation}
\frac{d}{d\tau}([1+\Phi(u)]\frac{du}{d\tau})=\frac{\partial_{v}[1+\Phi(u)]}{1+\Phi(u)}=0,\label{eq: t-z t+z error}
\end{equation}

which was integrated obtaining (eq. (3.28) of \cite{key-10})

\begin{equation}
\frac{du}{d\tau}=\frac{a}{1+\Phi},\label{eq: t+z error}
\end{equation}

while, using eq. (\ref{eq: t-z t+z}) we obtain

\begin{equation}
\frac{du}{d\tau}=\alpha,\label{eq: t-z}
\end{equation}

where $\alpha$ is an integration constant. 

From eqs. (\ref{eq: Ch3}) and (\ref{eq: t-z}), it is also 

\begin{equation}
\frac{dv}{d\tau}=\frac{\beta}{1+\Phi}\label{eq: t+z}
\end{equation}

where $\beta\equiv\frac{1}{\alpha}$, and

\begin{equation}
\tau=\beta u+\gamma,\label{eq: tau}
\end{equation}

where the integration constant $\gamma$ corresponds simply to the
retarded time coordinate translation $u$. Thus, without loss of generality,
it can be put equal to zero.

Instead, in \cite{key-10} the authors found (eq. (3.29) of \cite{key-10})
\begin{equation}
\frac{dv}{d\tau}=\frac{1}{a}\label{eq: t-z error}
\end{equation}

and (eq. (3.30) of \cite{key-10})
\begin{equation}
\tau=av+b.\label{eq: tau error}
\end{equation}

The difference between eqs. (\ref{eq: confronto 2}) and eqs. (\ref{eq: geodetiche Shibata})
generates the differences between the incorrect eqs. (3.28) and (3.29)
in \cite{key-10} and the correct eqs. (\ref{eq: t-z}) and (\ref{eq: t+z})
of this paper, i.e. in our work we obtain the conservation of $\frac{du}{d\tau}$
while the authors of \cite{key-10} obtained the conservation of $\frac{dv}{d\tau}.$ 

Now, let us see what is the meaning of the other integration constant
$\beta$ (see also \cite{key-10}). From eqs. (\ref{eq: t-z}) and
(\ref{eq: t+z}) the equation for $z$ can be written as

\begin{equation}
\frac{dz}{d\tau}=\frac{1}{2\beta}(\frac{\beta^{2}}{1+\Phi}-1).\label{eq: z}
\end{equation}

When it is $\Phi=0$ (i.e. before the SGW arrives at the test masses)
eq. (\ref{eq: z}) becomes
\begin{equation}
\frac{dz}{d\tau}=\frac{1}{2\beta}(\beta^{2}-1).\label{eq: z ad h nullo}
\end{equation}

But this is exactly the initial velocity of the test mass, thus $\beta=1$
has to be chosen because test masses are supposed at rest initially.
This also imply $\alpha=1$.

To find the motion of a test mass in the $z$ direction, we note that
from eq. (\ref{eq: tau}) it is $d\tau=du$, while from eq. (\ref{eq: t+z})
it is $dv=\frac{d\tau}{1+\Phi}$. 

Because it is also $z=\frac{v-u}{2},$ we obtain

\begin{equation}
dz=\frac{1}{2}(\frac{d\tau}{1+\Phi}-du),\label{eq: dz}
\end{equation}

which can be integrated as

\begin{equation}
\begin{array}{c}
z=z_{0}+\frac{1}{2}\int(\frac{du}{1+\Phi}-du)=\\
\\
=z_{0}-\frac{1}{2}\int_{-\infty}^{t-z}\frac{\Phi(u)}{1+\Phi(u)}du,
\end{array}\label{eq: moto lungo z}
\end{equation}

where $z_{0}$ is the initial position of the test mass. 

The displacement of the test mass in the $z$ direction can be written
as

\begin{equation}
\begin{array}{c}
\Delta z=z-z_{0}=-\frac{1}{2}\int_{-\infty}^{t-z_{0}-\Delta z}\frac{\Phi(u)}{1+\Phi(u)}du\\
\\
\simeq-\frac{1}{2}\int_{-\infty}^{t-z_{0}}\frac{\Phi(u)}{1+\Phi(u)}du.
\end{array}\label{eq: spostamento lungo z}
\end{equation}

The results can be also rewritten in function of the time coordinate
$t$:

\begin{equation}
\begin{array}{ccc}
x(t) & = & x_{0}\\
\\
y(t) & = & y_{0}\\
\\
z(t) & = & z_{0}-\frac{1}{2}\int_{-\infty}^{t-z_{0}}\frac{\Phi(u)}{1+\Phi(u)}d(u)\\
\\
\tau(t) & = & t-z(t),
\end{array}\label{eq: moto gauge Corda}
\end{equation}

which are different from 

\begin{equation}
\begin{array}{ccc}
x(t) & = & x_{0}\\
\\
y(t) & = & y_{0}\\
\\
z(t) & = & z_{0}+\frac{1}{2}I(t-z(t))\\
\\
\tau(t) & = & t+z(t)
\end{array}\label{eq: moto gauge Shibata}
\end{equation}

with 

\begin{equation}
I(t-z(t))\equiv\int_{-\infty}^{t-z_{0}}\Phi(u)du\label{eq: definizione di I}
\end{equation}

used from the authors of \cite{key-11} starting from the incorrect
geodesic equations (\ref{eq: geodetiche Shibata}).

In \cite{key-12}, for a regressive wave (i.e. in this case it is
$\Phi\equiv\Phi(t+z)$), one finds also:

\begin{equation}
\begin{array}{ccc}
\overline{x} & = & \overline{x}_{i}\\
\\
\overline{y} & = & \overline{y}_{i}\\
\\
\overline{z} & = & \overline{z}_{i}-\frac{1}{2}\int_{-\infty}^{t+\overline{z}_{i}}\delta\Phi(v)d(v)\\
\\
\frac{d\tau}{dv} & = & 1+\delta\Phi,
\end{array}\label{eq: moto gauge Shibata2}
\end{equation}

see eqs. 4.11 - 4.14, which are incorrect too (note: while in \cite{key-12}
the scalar field is indicated with $\delta\Phi$, and the coordinates
are barred, in this paper we use the same notations of \cite{key-12}
only in eq. (\ref{eq: moto gauge Shibata2}) and (\ref{eq: moto gauge Corda2})).
With an analysis analogous to the one used above, it is simple to
show that the correct equations of motion for a regressive SGW are

\begin{equation}
\begin{array}{ccc}
\overline{x} & = & \overline{x}_{i}\\
\\
\overline{y} & = & \overline{y}_{i}\\
\\
\overline{z} & = & \overline{z}_{i}+\frac{1}{2}\int_{-\infty}^{t+\overline{z}_{i}}\frac{\delta\Phi(v)}{1+\delta\Phi(v)}d(v)\\
\\
\tau & = & t+z,
\end{array}\label{eq: moto gauge Corda2}
\end{equation}

Now, let us resume what happens in the gauge (\ref{eq: gauge di Shibata}).
We have shown that in the $x-y$ plane an inertial test mass initially
at rest remains at rest throughout the entire passage of the SGW,
while in the $z$ direction an inertial test mass initially at rest
has a motion during the passage of the SGW. Thus, it could appear
that SGWs have a longitudinal effect and do not have a transversal
one (incorrect conclusion of \cite{key-10}), but the situation is
different as it will be shown in the following analysis.

\section{Analysis in the low frequencies approximation}

We have to clarify the use of words ``\emph{at rest}''. We want
to mean that the coordinates of test masses do not change in the presence
of the SGW in the $x-y$ plane, but we will show that the proper distance
between the beam-splitter and the mirror of an interferometer changes
even though their coordinates remain the same. On the other hand,
we will also show that the proper distance between the beam-splitter
and the mirror of an interferometer does not change in the $z$ direction
even if their coordinates change in the gauge (\ref{eq: metrica puramente scalare di Shibata}). 

A good way to analyse variations in the proper distance (time) is
by means of ``bouncing photons'': a photon can be launched from
the beam-splitter to be bounced back by the mirror (see ref. \cite{key-3,key-5,key-14}
and figure 1). 

\begin{figure}
\includegraphics{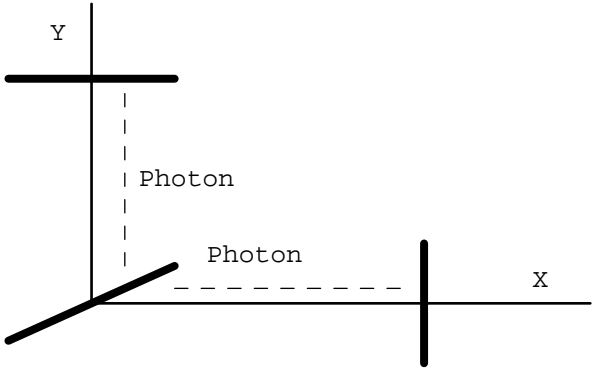}

\caption{Photons can be launched from the beam-splitter to be bounced back
by the mirror, adapted from ref. \cite{key-14}}
\end{figure}

In this section we only deal with the case in which the frequency
$f$ of the SGW is much smaller than $\frac{1}{T_{0}}=\frac{1}{L_{0}}$,
where $2T_{0}=2L_{0}$ is the total round-trip time of the photon
in absence of the SGW, exactly like in \cite{key-11}, but the correct
eqs. (\ref{eq: moto gauge Corda}) will be used differently from the
authors of \cite{key-11} that used the incorrect ones (\ref{eq: moto gauge Shibata}).
The analysis will be generalized to all frequencies in the next section.

Assuming that test masses are located along the $x$ axis and the
$z$ axis of the coordinate system the $y$ direction can be neglected
because the absence of the $y$ dependence in the metric (\ref{eq: metrica puramente scalare di Shibata})
implies that photon momentum in this direction is conserved \cite{key-3,key-5,key-14},
and the interval can be rewritten in the form

\begin{equation}
ds^{2}=[1+\Phi(t-z)](-dt^{2}+dx^{2}+dz^{2}).\label{eq: metrica + 4}
\end{equation}

Let us start by considering the interval for a photon which propagates
in the $x$ axis. Photon momentum in the $z$ direction is not conserved,
for the $z$ dependence in eq. (\ref{eq: metrica puramente scalare di Shibata})
\cite{key-3,key-5,key-14}. Thus, photons launched in the $x$ axis
will deflect out of this axis. But this effect can be neglected because
the photon deflection into the $z$ direction will be at most of order
$\Phi$ \cite{key-3,key-5,key-14}. Then, to first order in $\Phi$,
the $dz^{2}$ term can be neglected. Thus, from eq. (\ref{eq: metrica + 4})
one gets

\begin{equation}
ds^{2}=(1+\Phi)(-dt^{2})+(1+\Phi)dx^{2}.\label{eq: metrica + 3 lungo x}
\end{equation}

The condition for null geodesics ($ds^{2}=0$) for photons gives 

\begin{equation}
\frac{dx_{photon}}{dt}=\pm1\Rightarrow x_{photon}=const\pm t.\label{eq: null geodesic}
\end{equation}

In the gauge (\ref{eq: metrica puramente scalare di Shibata}) the
$x$ coordinates of the beam-splitter and the mirrors are unaffected
by the passage of the SGW (see the first of eqs. (\ref{eq: moto gauge Corda})),
then, from eq. (\ref{eq: null geodesic}) one gets that the interval,
in coordinate time $t$, that the photon takes for run one round trip
in the $x$ arm of the interferometer is

\begin{equation}
T=2L_{0}\label{eq:  tempo doppio}
\end{equation}

(i.e. the photon leaves the beam-splitter at $t=0$ and returns a
$t=T$). But this quantity is not invariant under coordinate transformations
\cite{key-3,key-5,key-14}, and we have to work in terms of the beam-splitter
proper time which measures the physical length of the arms. In this
way, we call $\tau(t)$ and $z_{b}(t)$ the proper time and $z$ coordinate
of the beam-splitter at time coordinate $t$ with initial condition
$z_{b}(-\infty)=0$. From eqs. (\ref{eq: moto gauge Corda}) it is

\begin{equation}
\begin{array}{ccc}
z_{b}(t) & = & -\frac{1}{2}\int_{-\infty}^{t-z_{b}(t)}\frac{\Phi(u)}{1+\Phi(u)}du\\
\\
\tau(t) & =t & +\frac{1}{2}\int_{-\infty}^{t-z_{b}(t)}\frac{\Phi(u)}{1+\Phi(u)}du.
\end{array}\label{eq: bm lungo z}
\end{equation}

Thus, calling $\tau_{x}$ the proper time interval that the photon
takes to run a round-trip in the $x$ arm, it is

\begin{equation}
\begin{array}{c}
\tau_{x}=\tau(T)-\tau(0)=T+\frac{1}{2}\int_{-z_{b}(0)}^{t-z_{b}(t)}\frac{\Phi(u)}{1+\Phi(u)}du\simeq\\
\\
\simeq T+\frac{1}{2}\Phi_{0}[T+z_{b}(0)-z_{b}(T)]\simeq\\
\\
\simeq2L_{0}(1+\frac{1}{2}\Phi_{0}),
\end{array}\label{eq: tempo proprio lungo x}
\end{equation}

which is the same result obtained in \cite{key-11}, but here it is
obtained from the correct equations of motion. 

In the above computation, eq. (\ref{eq:  tempo doppio}) have been
used and, by considering only the first order in $\Phi,$ with $\Phi\ll1$,
the scalar field $\Phi$ has also been considered ``frozen'' at
a fixed value $\Phi_{0}$. Note that $z_{b}(0)-z_{b}(T)$ is second
order in $\Phi_{0}$.

The computation of \cite{key-11} is correct but it starts from incorrect
equations of motion, i.e. the authors of \cite{key-11} casually obtained
the correct result (\ref{eq: tempo proprio lungo x}) starting from
incorrect equations of motion. This is because the correct equation
of motion (\ref{eq: moto gauge Corda2}) for a regressive SGW are
casually very similar to the wrong ones (\ref{eq: moto gauge Shibata})
for a progressive SGW. In fact, using the notation of \cite{key-11}
and rewriting the correct equations of motion for a regressive wave
(\ref{eq: moto gauge Corda2}), one gets

\begin{equation}
\begin{array}{ccc}
x(t) & = & x_{0}\\
\\
y(t) & = & y_{0}\\
\\
z(t) & = & z_{0}+\frac{1}{2}\int_{-\infty}^{t+z_{0}}\frac{\Phi(u)}{1+\Phi(u)}d(u)\\
\\
\tau(t) & = & t+z(t),
\end{array}\label{eq: moto gauge Corda 3}
\end{equation}

and, because it is $\Phi(u)\ll1$ one obtains

\begin{equation}
\frac{\Phi(u)}{1+\Phi(u)}\simeq\Phi(u),\label{eq: circa}
\end{equation}

and the only difference between eqs. (\ref{eq: moto gauge Corda 3})
and eqs. (\ref{eq: moto gauge Shibata}) is the different parametrization
of the wave: regressive in eq. (\ref{eq: moto gauge Corda 3}), progressive
in eq. (\ref{eq: moto gauge Shibata}).

Now, let us consider the $z$ direction: the $x$ direction can be
neglected because the absence of the $x$ dependence in the metric
(\ref{eq: metrica + 4}) implies that photon momentum in this direction
is conserved \cite{key-3,key-5,key-14}. From eq. (\ref{eq: metrica + 4})
it is:

\begin{equation}
ds^{2}=(1+\Phi)(-dt^{2})+(1+\Phi)dz^{2},\label{eq: metrica puramente piu di Corda lungo z}
\end{equation}

and the condition for null geodesics ($ds^{2}=0$) for photons gives 

\begin{equation}
\frac{dz_{photon}}{dt}=\pm1\Rightarrow z_{photon}=const\pm t.\label{eq: null geodesic2}
\end{equation}
We suppose that the photon leaves the beam splitter at $t=0$; let
us ask: how much time does the photon need to arrive at the mirror
in the $z$ axis? Calling $T_{1}$ this time one needs the condition

\begin{equation}
z_{b}(0)+T_{1}=z_{m}(T_{1}),\label{eq: z.m}
\end{equation}

where $z_{m}(t)$ is the $z$ coordinate of the mirror in the $z$
axis at coordinate time $t$ with $z_{m}(-\infty)=L_{0}$. In the
same way, when returning from the mirror, the photon arrives again
at the beam-splitter at $t=T_{z}=T_{1}+T_{2}$, then

\begin{equation}
z_{m}(T_{1})-T_{2}=z_{b}(T_{z}).\label{eq: z.m2}
\end{equation}

Subtracting eq. (\ref{eq: z.m2}) from eq. (\ref{eq: z.m}) we get

\begin{equation}
T_{z}=T_{1}+T_{2}=[z_{m}(T_{1})-z_{b}(0)]+[z_{m}(T_{1})-z_{b}(T_{z})].\label{eq: tempo}
\end{equation}

From eq. (\ref{eq: moto gauge Corda}) the equations of motion for
$z_{b}$ and $z_{m}$ are:

\begin{equation}
\begin{array}{ccc}
z_{m}(t) & = & L_{0}-\frac{1}{2}\int_{-\infty}^{t-z_{m}(t)}\frac{\Phi(u)}{1+\Phi(u)}du\\
\\
z_{b}(t) & = & -\frac{1}{2}\int_{-\infty}^{t-z_{b}(t)}\frac{\Phi(u)}{1+\Phi(u)}du,
\end{array}\label{eq: bm lungo z bis}
\end{equation}

and, substituting them in eq. (\ref{eq: tempo}), one obtains

\begin{equation}
T_{z}=2L_{0}-\frac{1}{2}\int_{-z_{b}(0)}^{T_{1}-z_{m}(T_{1})}\frac{\Phi(u)}{1+\Phi(u)}du-\frac{1}{2}\int_{T_{z}-z_{b}(T_{z})}^{T_{1}-z_{m}(T_{1})}\frac{\Phi(u)}{1+\Phi(u)}du.\label{eq: tempo lungo z}
\end{equation}

From eq. (\ref{eq: z.m}), the first integral in eq. (\ref{eq: tempo lungo z})
is zero. The second integral is simple to compute by considering the
SGW frozen at a value $\Phi_{0}$. To first order in this value we
get

\begin{equation}
\begin{array}{c}
-\frac{1}{2}\int_{T_{z}-z_{b}(T_{z})}^{T_{1}-z_{m}(T_{1})}\frac{\Phi(u)}{1+\Phi(u)}du\simeq-\frac{1}{2}\Phi_{0}[T_{1}-z_{m}(T_{1})-T_{z}+z_{b}(T_{z})]\simeq\\
\\
\simeq-\frac{1}{2}\Phi_{0}(L_{0}-L_{0}-2L_{0})=+\frac{1}{2}\Phi_{0}2L_{0}.
\end{array}\label{eq: int2}
\end{equation}

In this way, eq. (\ref{eq: tempo lungo z}) becomes

\begin{equation}
T_{z}=(1+\frac{1}{2}\Phi_{0})2L_{0}.\label{eq: tempo lungo z2}
\end{equation}

Then, calling $\tau_{z}$ the proper time interval that the photon
takes to run a round-trip in the $z$ arm, with the same kind of analysis
which leaded to eq. (\ref{eq: tempo proprio lungo x}), we get

\begin{equation}
\begin{array}{c}
\tau_{z}=\tau(T_{z})-\tau(0)=T_{z}+\frac{1}{2}\int_{-z_{b}(0)}^{t-z_{b}(t)}\frac{\Phi(u)}{1+\Phi(u)}du\simeq\\
\\
\simeq T_{z}+\frac{1}{2}\Phi_{0}[T_{z}+z_{b}(0)-z_{b}(T)]\simeq\\
\\
\simeq T_{z}(1-\frac{1}{2}\Phi_{0})\simeq2L_{0},
\end{array}\label{eq: tempo proprio lungo z 3}
\end{equation}

which is also the same result of the correspondent equation in \cite{key-11}:

\begin{equation}
\tau_{z}\simeq T_{z}(1+\frac{1}{2}\Phi_{0})\simeq2L_{0}.\label{eq: MN2}
\end{equation}
 This is also due to the similarity of eqs. (\ref{eq: moto gauge Corda 3})
and eqs. (\ref{eq: moto gauge Shibata}): the difference in sign between
eqs. (\ref{eq: tempo proprio lungo z 3}) and (\ref{eq: MN2}) is
due to the difference between the progressive and the regressive wave.
In \cite{key-11} we find also 

\begin{equation}
T_{z}=(1-\frac{1}{2}\Phi_{0})2L_{0},\label{eq: MN3}
\end{equation}

i.e. a difference in sign with eq. (\ref{eq: tempo lungo z2}), which
compensates the difference in sign between eqs. (\ref{eq: tempo proprio lungo z 3})
and (\ref{eq: MN2}).

Thus, from eqs. (\ref{eq: tempo proprio lungo x}) and (\ref{eq: tempo proprio lungo z 3})
one can say that there is a variation of the proper distance in the
$x$ direction (transverse effect of the SGW), while there is not
a variation of the proper distance in the $z$ direction (no longitudinal
effect).

\section{Generalized analysis}

Now, let us generalize to all the frequencies the previous result,
with an analysis that, with a transform of the time coordinate to
the proper time, generalizes to the gauge (\ref{eq: gauge di Shibata})
the analysis of \cite{key-14}. In this way the response function
of the interferometer to SGWs in the gauge (\ref{eq: gauge di Shibata})
will be obtained.

Let us start with the $x$ arm of the interferometer. The condition
of null geodesic (\ref{eq: null geodesic}) can be also rewritten
as

\begin{equation}
dt^{2}=dx^{2}.\label{eq: metrica puramente piu' di Corda lungo x 2}
\end{equation}

In \cite{key-3} we used the condition of null geodesic in the case
of the transverse-traceless (TT) gauge for tensor waves \cite{key-18}
to obtain the coordinate velocity of the photon which was used for
calculations of the photon propagation times between the test masses
(eq. (7) in \cite{key-3}). But from eq. (\ref{eq: metrica puramente piu' di Corda lungo x 2})
we see that the coordinate velocity of the photon in the gauge (\ref{eq: gauge di Shibata})
is equal to the speed of light. Thus, in this case, the analysis of
\cite{key-3} cannot be used starting directly from the condition
of null geodesic. Let us ask which is the important difference between
the gauge (\ref{eq: gauge di Shibata}) and the TT gauge for tensor
waves analysed in \cite{key-3}. The answer is that the TT gauge of
\cite{key-3} is a ``synchrony gauge'', a coordinate system in which
the time coordinate $t$ is exactly the proper time (about the synchrony
coordinate system see Cap. (9) of ref. \cite{key-13}). In the coordinates
(\ref{eq: metrica puramente scalare di Shibata}) $t$ is only a time
coordinate. The rate $d\tau$ of the proper time is related to the
rate $dt$ of the time coordinate from \cite{key-13}

\begin{equation}
d\tau^{2}=g_{00}dt^{2}.\label{eq: relazione temporale}
\end{equation}

Only making the time transform (\ref{eq: relazione temporale}) the
analysis of \cite{key-3} can be applied to the gauge (\ref{eq: gauge di Shibata}).

From eq. (\ref{eq: metrica + 3 lungo x}) we get $g_{00}=(1+\Phi)$.
Then, using eq. (\ref{eq: metrica puramente piu' di Corda lungo x 2}),
one obtains

\begin{equation}
d\tau^{2}=(1+\Phi)dx^{2},\label{eq: relazione spazial-temporale}
\end{equation}

which gives 

\begin{equation}
d\tau=\pm(1+\Phi)^{\frac{1}{2}}dx.\label{eq: relazione temporale 2}
\end{equation}

Now, we show that the analysis of \cite{key-3,key-5,key-14} works
in this case too. 

From eqs. (\ref{eq: moto gauge Corda}) the coordinates of the beam-splitter
$x_{b}=l$ and of the mirror $x_{m}=l+L_{0}$ do not changes under
the influence of the SGW in the gauge (\ref{eq: metrica puramente scalare di Shibata}).
Hence, the proper duration of the forward trip can be found as

\begin{equation}
\tau_{1}(t)=\int_{l}^{L_{0}+l}[1+\Phi(t)]^{\frac{1}{2}}dx.\label{eq: relazione temporale 2 bis}
\end{equation}

To first order in $\Phi$ this integral can be approximated with

\begin{equation}
\tau_{1}(t)=T_{0}+\frac{1}{2}\int_{l}^{L_{0}+l}\Phi(t')dx\label{eq: durata volo andata approssimata in Corda x}
\end{equation}

where

\begin{center}
$t'=t-(l+L_{0}-x)$.
\par\end{center}

In the last equation $t'$ is the delay time (i.e. $t$ is the time
at which the photon arrives in the position $l+L_{0}$, thus $l+L_{0}-x=t-t'$
\cite{key-3,key-5,key-14}).

In the same way, for the proper duration of the return trip, we write

\begin{equation}
\tau_{2}(t)=T_{0}+\frac{1}{2}\int_{l+L_{0}}^{l}\Phi(t')(-dx)\label{eq: durata volo ritorno approssimata in Corda x}
\end{equation}

where now 

\begin{center}
$t'=t-(x-l)$
\par\end{center}

is the delay time and

\begin{center}
$T_{0}=L_{0}$ 
\par\end{center}

is the transit proper time of the photon in the absence of the SGW,
which also corresponds to the transit coordinate time of the photon
in the presence of the SGW (see eq. (\ref{eq: metrica puramente piu' di Corda lungo x 2})).

Thus, the round-trip proper time will be the sum of $\tau_{2}(t)$
and $\tau_{1}(t-T_{0})$. Then, to first order in $\Phi$, the proper
duration of the round-trip will be

\begin{equation}
\tau_{r.t.}(t)=\tau_{1}(t-T_{0})+\tau_{2}(t).\label{eq: durata round trip}
\end{equation}

By using eqs. (\ref{eq: durata volo andata approssimata in Corda x})
and (\ref{eq: durata volo ritorno approssimata in Corda x}) one immediately
gets that deviations of this round-trip proper time (i.e. proper distance)
from its unperturbed value are given by

\begin{equation}
\delta\tau(t)=\frac{1}{2}\int_{l}^{L_{0}+l}[\Phi(t-2T_{0}+x-l)+\Phi(t-x+l)]dx.\label{eq: variazione temporale in gauge comovente}
\end{equation}

Eq. (\ref{eq: variazione temporale in gauge comovente}) generalizes
eq. (\ref{eq: tempo proprio lungo x}) which was derived in the low
frequencies approximation. Defining the signal in the arm in the $x$
axis like

\begin{equation}
\frac{\delta\tau(t)}{T_{0}}\equiv\frac{1}{2T_{0}}\int_{l}^{L_{0}+l}[\Phi(t-2T_{0}+x-l)+\Phi(t-x+l)]dx,\label{eq: signal}
\end{equation}

the analysis can be translated in the frequency domain using the Fourier
transform of the field which is \cite{key-14}

\begin{equation}
\tilde{\Phi}(\omega)=\int_{-\infty}^{\infty}dt\Phi(t)\exp(i\omega t).\label{eq: trasformata di fourier}
\end{equation}

By using definition (\ref{eq: trasformata di fourier}), from eq.
(\ref{eq: signal}) we get

\begin{equation}
\frac{\delta\tilde{\tau}(\omega)}{T_{0}}=\Upsilon(\omega)\tilde{\Phi}(\omega),\label{eq: fourier in gage comoventi e Corda}
\end{equation}

where $\Upsilon(\omega)$ is the response of the $x$ arm of the interferometer
to SGWs \cite{key-14}:

\begin{equation}
\Upsilon(\omega)=\frac{\exp(2i\omega T_{0})-1}{2i\omega T_{0}}.\label{eq: risposta in gages comovente e Corda}
\end{equation}

Notice that an analysis similar to the one performed in this Section
has been used in ref. \cite{key-3} for tensor waves.

Now, let us see what happens in the $z$ coordinate (see figure 2).
\begin{figure}
\includegraphics{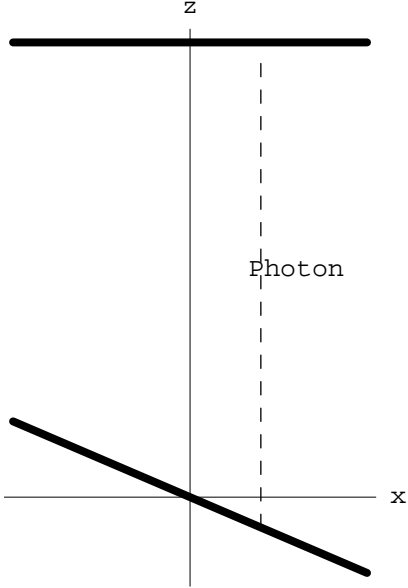}

\caption{the beam splitter and the mirror are located in the direction of the
incoming GW, adapted from ref. \cite{key-14}}
\end{figure}

From eq. (\ref{eq: metrica puramente piu di Corda lungo z}) and the
condition $ds^{2}=0$ for null geodesics we get

\begin{equation}
dz=\pm dt.\label{eq: metrica puramente piu di Corda lungo z 2}
\end{equation}

But, from the last of eqs. (\ref{eq: moto gauge Corda}) the proper
time is

\begin{equation}
d\tau(t)=dt-dz,\label{eq: tempo proprio lungo z in Corda}
\end{equation}

and, combining eq. (\ref{eq: metrica puramente piu di Corda lungo z 2})
with eq. (\ref{eq: tempo proprio lungo z in Corda}), one obtains

\begin{equation}
d\tau(t)=dt\mp dt.\label{eq: tempo proprio lungo z in Corda 2}
\end{equation}

Hence 

\begin{equation}
\tau_{1}(t)=0\label{eq:  tempo di propagazione andata gauge Corda lungo z}
\end{equation}

for the forward trip

and

\begin{equation}
\tau_{2}(t)=\int_{0}^{T_{0}}2dt=2T_{0}\label{eq:  tempo di propagazione ritorno gauge Corda  lungo z}
\end{equation}

for the return trip. At the end we get

\begin{equation}
\tau(t)=\tau_{1}(t)+\tau_{2}(t)=2T_{0}.\label{eq: tempo proprio totale lungo z in Corda}
\end{equation}

Thus, it is $\delta\tau=\delta L_{0}=0$, i.e. there is no longitudinal
effect. This is a direct consequence of the fact that a SGW propagates
at the speed of light. In this way in the forward trip the photon
travels at the same speed of the SGW and its proper time is equal
to zero (eq. (\ref{eq:  tempo di propagazione andata gauge Corda lungo z})),
while in the return trip the photon travels \textbf{against} the SGW
and its proper time redoubles (eq. (\ref{eq:  tempo di propagazione ritorno gauge Corda  lungo z})).

\section{The detector pattern}

As the arms of an interferometer are in general in the $\overrightarrow{u}$
and $\overrightarrow{v}$ directions, to compute the line element
in the $\overrightarrow{u}$ and $\overrightarrow{v}$ directions,
a spatial rotation of the coordinates (\ref{eq: metrica puramente scalare di Shibata})
is needed:

\begin{equation}
\begin{array}{ccc}
u & = & -x\cos\theta\cos\phi+y\sin\phi+z\sin\theta\cos\phi\\
\\
v & = & -x\cos\theta\sin\phi-y\cos\phi+z\sin\theta\sin\phi\\
\\
w & = & x\sin\theta+z\cos\theta,
\end{array}\label{eq: rotazione}
\end{equation}

or, in terms of the $x,y,z$ frame:

\begin{equation}
\begin{array}{ccc}
x & = & -u\cos\theta\cos\phi-v\cos\theta\sin\phi+w\sin\theta\\
\\
y & = & u\sin\phi-v\cos\phi\\
\\
z & = & u\sin\theta\cos\phi+v\sin\theta\sin\phi+w\cos\theta.
\end{array}\label{eq: rotazione 2}
\end{equation}

In this way, the SGW is propagating from an arbitrary direction $\overrightarrow{r}$
to the interferometer (see figure 3). 

\begin{figure}
\includegraphics{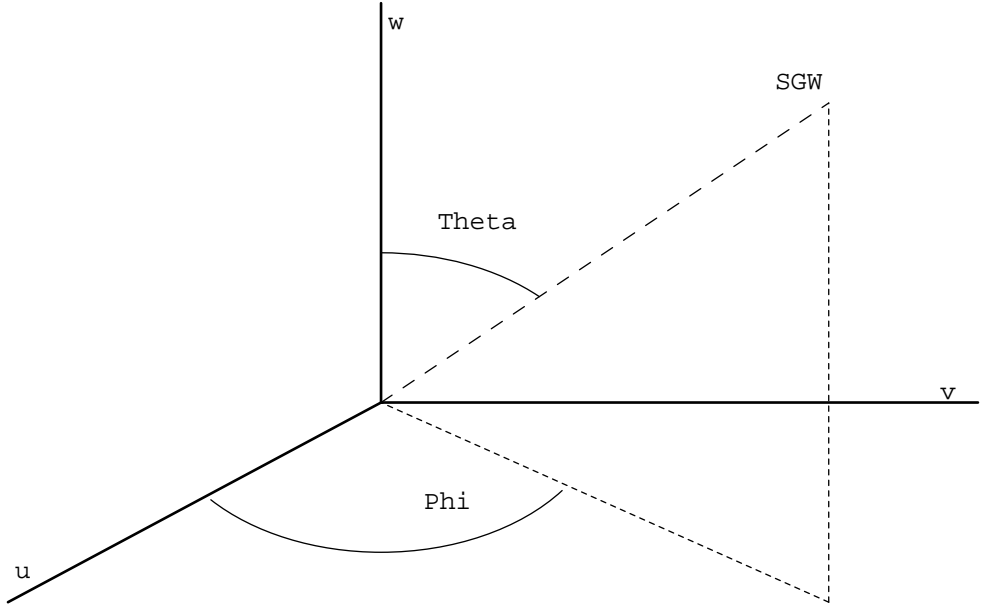}

\caption{a SGW incoming from an arbitrary direction, adapted from ref. \cite{key-14}}
\end{figure}
The metric tensor transforms like (see Chap. (10) of ref. \cite{key-13}): 

\begin{equation}
g^{ik}=\frac{\partial x^{i}}{\partial x'^{l}}\frac{\partial x^{k}}{\partial x'^{m}}g'^{lm}.\label{eq: trasformazione metrica}
\end{equation}

Using eq. (\ref{eq: rotazione}), eq. (\ref{eq: rotazione 2}) and
eq. (\ref{eq: trasformazione metrica}), in the new rotated frame,
the line element (\ref{eq: metrica puramente scalare di Shibata})
in the $\overrightarrow{u}$ direction becomes (here one can neglect
the $v$ and $w$ directions because bouncing photons will be used
and the photon deflection into the $v$ and $w$ directions will be
at most of order $\Phi,$ then, to first order in $\Phi$, the $dv^{2}$
and $dw^{2}$ terms can be neglected \cite{key-14}):

\begin{equation}
ds^{2}=[1+(1-\sin^{2}\theta\cos^{2}\phi)\Phi(t-u\sin\theta\cos\phi)](du^{2}-dt^{2}).\label{eq: metrica scalare lungo u 2}
\end{equation}

Considering a photon launched from the beam-splitter to be bounced
back by the mirror, the condition for null geodesics ($ds^{2}=0$)
in eq. (\ref{eq: metrica scalare lungo u 2}) gives 

\begin{equation}
du^{2}=dt^{2}.\label{eq:u uguale t}
\end{equation}

Thus, also in this case, the analysis of \cite{key-3} cannot be used
starting directly from the condition of null geodesic. But a generalization
of previous analysis can be used. In fact, also the metric (\ref{eq: metrica scalare lungo u 2})
is not a ``synchrony coordinate system'', thus, also in this line
element, $t$ is only a time coordinate. The rate $d\tau$ of the
proper time is related to the rate $dt$ of the time coordinate from
eq. (\ref{eq: relazione temporale}).

From eq. (\ref{eq: metrica scalare lungo u 2}) we get

\begin{equation}
g_{00}=[1+(1-\sin^{2}\theta\cos^{2}\phi)\Phi(t-u\sin\theta\cos\phi)].\label{eq: g00}
\end{equation}
. 

Then, using eq. (\ref{eq:u uguale t}), we obtain
\begin{equation}
d\tau^{2}=[1+(1-\sin^{2}\theta\cos^{2}\phi)\Phi(t-u\sin\theta\cos\phi)]du^{2}\label{eq: relazione spazio tempo}
\end{equation}

which gives 
\begin{equation}
d\tau=\pm[1+(1-\sin^{2}\theta\cos^{2}\phi)\Phi(t-u\sin\theta\cos\phi)]{}^{\frac{1}{2}}du.\label{eq: relazione temporale 2 tris}
\end{equation}

We put the beam splitter in the origin of the new coordinate system
(i.e. $u_{b}=0$, $v_{b}=0$, $w_{b}=0$). From eqs. (\ref{eq: moto gauge Corda})
one gets that an inertial test mass initially at rest in the $x-y$
plane of the coordinates (\ref{eq: metrica puramente scalare di Shibata}),
remains at rest throughout the entire passage of the SGW. We also
know that the coordinates of the beam-splitter and of the mirror change
under the influence of the SGW in the $z$ direction, but this fact
does not influence the total variation of the round trip proper time
of the photon (eq. (\ref{eq: tempo proprio totale lungo z in Corda})).
Then, in the computation of the variation of the proper distance in
the gauge (\ref{eq: metrica puramente scalare di Shibata}), the coordinates
of the beam-splitter $u_{b}=0$ and of the mirror $u_{m}=L_{0}$ can
be considered fixed even in the $u-v$ plane, because the rotation
(\ref{eq: rotazione}) does not change the situation. Thus, the proper
duration of the forward trip can be found as

\begin{equation}
\tau_{1}(t)=\int_{0}^{L_{0}}[1+(1-\sin^{2}\theta\cos^{2}\phi)\Phi(t-u\sin\theta\cos\phi)]{}^{\frac{1}{2}}du\label{eq: relazione spazio tempo 2}
\end{equation}

with 

\begin{center}
$t'=t-(L_{0}-u)$.
\par\end{center}

In the last equation $t'$ is the delay time (see Section 4).

To first order in $\Phi$ this integral is well approximated with

\begin{equation}
\tau_{1}(t)=T_{0}+\frac{1-\sin^{2}\theta\cos^{2}\phi}{2}\int_{0}^{L_{0}}\Phi(t'-u\sin\theta\cos\phi)du,\label{eq: durata volo andata approssimata u}
\end{equation}

where

\begin{center}
$T_{0}=L_{0}$ 
\par\end{center}

is the transit time of the photon in the absence of the SGW. Similarly,
the duration of the return trip will be
\begin{equation}
\tau_{2}(t)=T_{0}+\frac{1-\sin^{2}\theta\cos^{2}\phi}{2}\int_{L_{0}}^{0}\Phi(t'-u\sin\theta\cos\phi)(-du),\label{eq: durata volo ritorno approssimata u}
\end{equation}

though now the delay time is 

\begin{center}
$t'=t-(u-l)$.
\par\end{center}

The round-trip time will be the sum of $\tau_{2}(t)$ and $\tau_{1}[t-T_{0}]$.
Thus, to first order in $\Phi$, the proper duration of the round-trip
will be

\begin{equation}
\tau_{r.t.}(t)=\tau_{1}[t-T_{0}]+\tau_{2}(t).\label{eq: durata round trip bis}
\end{equation}

Using eqs. (\ref{eq: durata volo andata approssimata u}) and (\ref{eq: durata volo ritorno approssimata u})
one immediately gets that deviations of this round-trip time (i.e.
proper distance) from its unperturbed value are given by

\begin{equation}
\begin{array}{c}
\delta\tau(t)=\frac{1-\sin^{2}\theta\cos^{2}\phi}{2}\int_{0}^{L_{0}}[\Phi(t-2L_{0}+u(1-\sin\theta\cos\phi))+\\
\\
+\Phi(t-u(1+\sin\theta\cos\phi))]du.
\end{array}\label{eq: variazione temporale in u}
\end{equation}

By using the Fourier transform of the scalar field defined by eq.
(\ref{eq: trasformata di fourier}), in the frequency domain it is:

\begin{equation}
\delta\tilde{\tau}(\omega)=(1-\sin^{2}\theta\cos^{2}\phi)\tilde{H}_{u}(\omega,\theta,\phi)\Phi(\omega)\label{eq: segnale in frequenza lungo u}
\end{equation}

where

\begin{equation}
\begin{array}{c}
\tilde{H}_{u}(\omega,\theta,\phi)=\frac{-1+\exp(2i\omega L_{0})}{2i\omega(1-\sin^{2}\theta\cos^{2}\phi)}+\\
\\
+\frac{\sin\theta\cos\phi((1+\exp(2i\omega L_{0})-2\exp i\omega L_{0}(1+\sin\theta\cos\phi)))}{2i\omega(1-\sin^{2}\theta\cos^{2}\phi)},
\end{array}\label{eq: fefinizione Hu}
\end{equation}

and one immediately obtains that $\tilde{H}_{u}(\omega,\theta,\phi)\rightarrow L_{0}$
when $\omega\rightarrow0$.

Thus, the total response function of the arm of the interferometer
in the $\overrightarrow{u}$ direction to the SGW is:

\begin{equation}
\Upsilon_{u}^{SGW}(\omega)=\frac{1-\sin^{2}\theta\cos^{2}\phi}{L_{0}}\tilde{H}_{u}(\omega,\theta,\phi).\label{eq: risposta totale lungo u 2}
\end{equation}

In the same way, the line element (\ref{eq: metrica puramente scalare di Shibata})
in the $\overrightarrow{v}$ direction becomes:

\begin{equation}
ds^{2}=[1+(1-\sin^{2}\theta\sin^{2}\phi-)\Phi(t-v\sin\theta\sin\phi)](dv^{2}-dt^{2}),\label{eq: metrica + lungo v}
\end{equation}

and, with the same kind of analysis used for the $\overrightarrow{u}$
direction, the response function of the $\overrightarrow{v}$ arm
of the interferometer to the SGW results:

\begin{equation}
\Upsilon_{v}^{SGW}=\frac{1-\sin^{2}\theta\sin^{2}\phi}{L_{0}}\tilde{H}_{v}(\omega,\theta,\phi),\label{eq: risposta totale lungo v}
\end{equation}

where

\begin{equation}
\begin{array}{c}
\tilde{H}_{v}(\omega,\theta,\phi)=\frac{-1+\exp(2i\omega L_{0})}{2i\omega(1-\sin^{2}\theta\sin^{2}\phi)}+\\
\\
+\frac{\sin\theta\sin\phi((1+\exp(2i\omega L_{0})-2\exp i\omega L(1+\sin\theta\sin\phi)))}{2i\omega(1-\sin^{2}\theta\sin^{2}\phi)},
\end{array}\label{eq: fefinizione Hv}
\end{equation}

and we see that also $\tilde{H}_{v}(\omega,\theta,\phi)\rightarrow L_{0}$
when $\omega\rightarrow0$.

Then, the detector pattern of an interferometer to the SGW is:

\begin{equation}
\begin{array}{c}
\tilde{H}^{SGW}(\omega)=\frac{1-\sin^{2}\theta\cos^{2}\phi}{L_{0}}\tilde{H}_{u}(\omega,\theta,\phi)-\frac{1-\sin^{2}\theta\sin^{2}\phi}{L_{0}}\tilde{H}_{v}(\omega,\theta,\phi)=\\
\\
=\frac{\sin\theta}{2i\omega L}\{\cos\phi[1+\exp(2i\omega L)-2\exp i\omega L(1+\sin\theta\cos\phi)]+\\
\\
-\sin\phi[1+\exp(2i\omega L_{0})-2\exp i\omega L_{0}(\sin\theta\sin\phi-1)]\},
\end{array}\label{eq: risposta totale Virgo}
\end{equation}

that is exactly the detector pattern of eq. (150) of \cite{key-14},
where the computation was made in the TT gauge, and the same result
of eq. (18) of \cite{key-19}, where this response function has been
used to analyse the cross-correlation between the Virgo interferometer
and the MiniGRAIL resonant sphere for the detection of SGWs.

For a sake of clearness, the derivation of this detector pattern in
the TT gauge will be sketch in the appendix \cite{key-17}. 

In the low frequencies limit ($\omega\rightarrow0$) eq. (\ref{eq: risposta totale Virgo})
is also in perfect agreement with the detector pattern of eq. (15)
of \cite{key-11}, and also with the low frequencies detector pattern
of \cite{key-15,key-16}:

\begin{equation}
\tilde{H}^{SGW}(\omega\rightarrow0)=-\sin^{2}\theta\cos2\phi.\label{eq: risposta totale approssimata}
\end{equation}

The detector pattern of eq. (\ref{eq: risposta totale Virgo}) is
different from the one in eq. (5.25) of \cite{key-12}, because in
\cite{key-12} the computation was made starting from incorrect equations
of motion. The similarity between the two detector patterns is due
to the fact that the correct equations of motion (\ref{eq: moto gauge Corda})
for a progressive SGW are casually very similar to the incorrect ones
(\ref{eq: moto gauge Shibata2}) for a regressive SGW used in \cite{key-12}
(see also comments in Section 3 about the casual similarity between
wrong and correct equations of motion).

Notice that also in this case an analysis similar to the one performed
in this Section has been used in ref. \cite{key-3} for tensor waves.

The absolute value of the total response function of the Virgo interferometer
($L=3$ Km) for SGWs with $\theta=\frac{\pi}{4}$ and $\phi=\frac{\pi}{3}$
and the angular dependence of the response of the Virgo interferometer
for a SGW with a frequency of $f=100Hz$ are respectively shown in
figs. 4 and 5.

\begin{figure}
\includegraphics{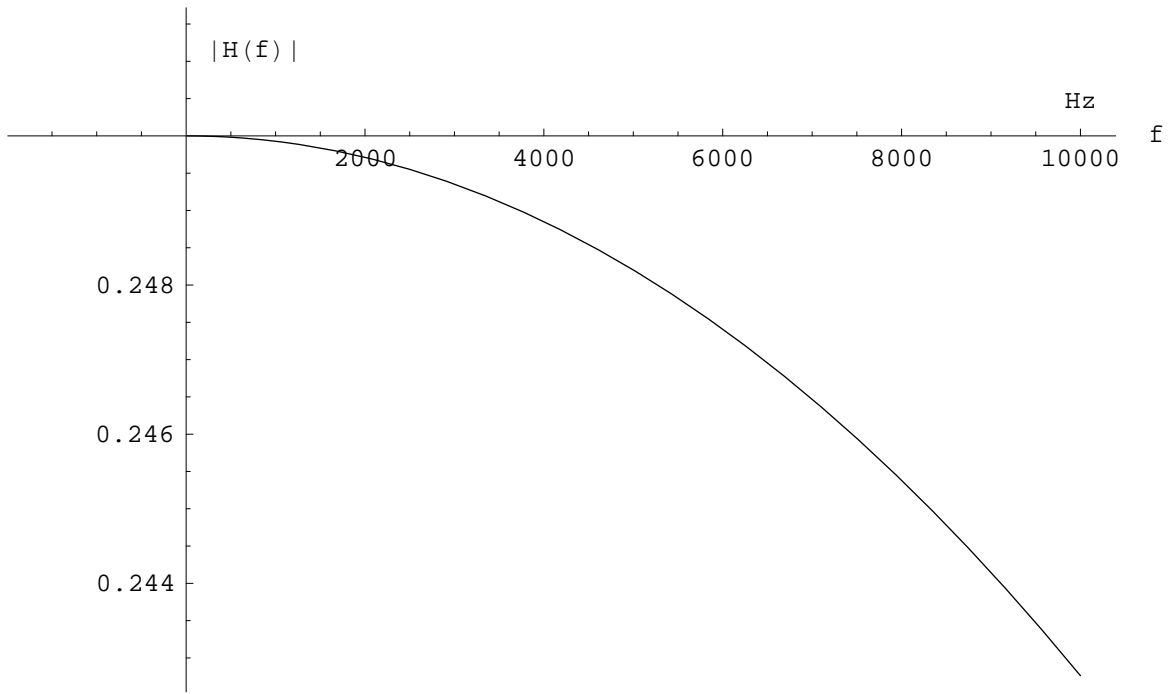}

\caption{the absolute value of the total response function of the Virgo interferometer
to SGWs for $\theta=\frac{\pi}{4}$ and $\phi=\frac{\pi}{3}$, adapted
from ref. \cite{key-14} }
\end{figure}
\begin{figure}
\includegraphics{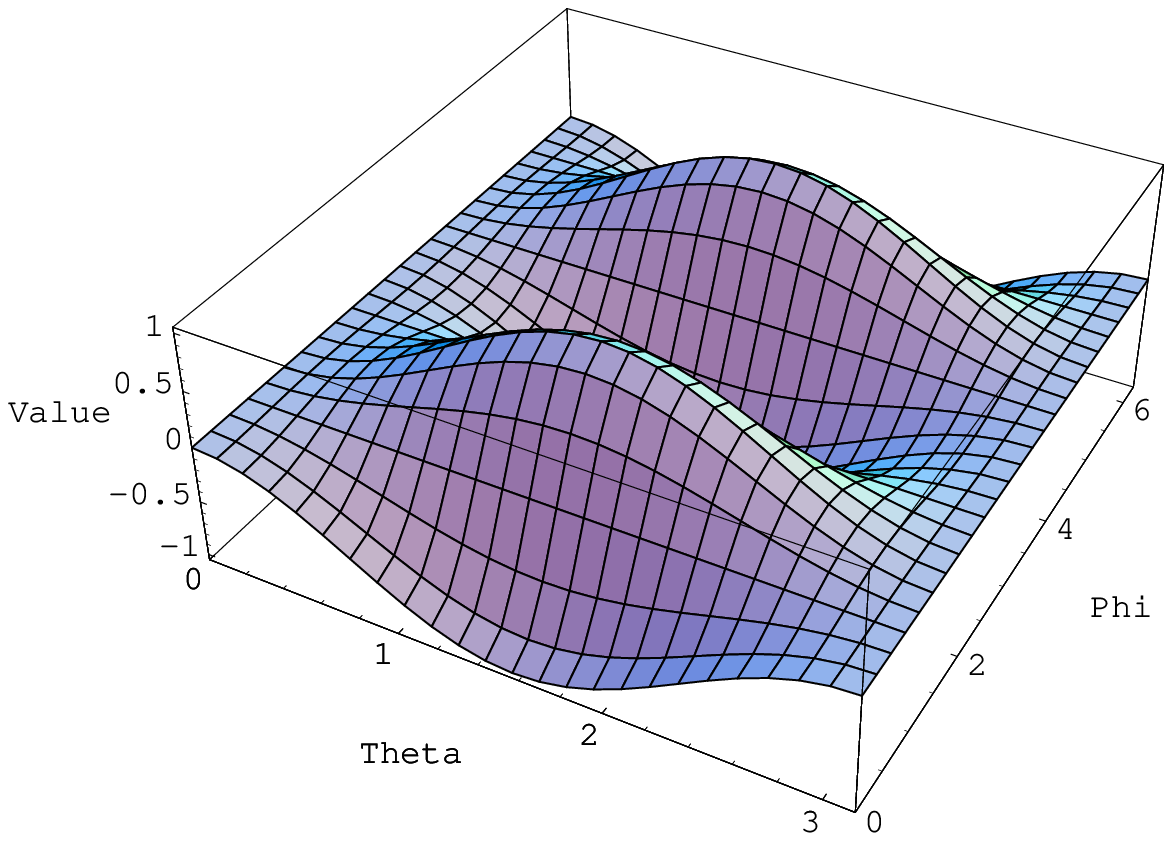}

\caption{the angular dependence of the response of the Virgo interferometer
for a SGW with a frequency of $f=100Hz$, adapted from ref. \cite{key-14}}
\end{figure}

\section{Conclusions }

Following some ideas in the Landau\textquoteright{}s book, some corrections
about errors in the old literature on SGWs have been released and
discussed. In the analysis of the response of interferometers the
computation has been first performed in the low frequencies approximation.
After this, the analysis has been applied to all SGWs in the full
frequency and angular dependences. 

The presented results are in agreement with the more recent literature
on SGWs.

\section*{Acknowledgements}

The R. M. Santilli Foundation has to be thanked for partially supporting
this letter (Research Grant of the R. M. Santilli Foundation Number
RMS-TH-5735A2310). I thank \foreignlanguage{italian}{Maria Felicia
De Laurentis for useful discussions on gravity waves. I also thanks
an unknown referee for useful comments.}

\section*{Appendix}

For a sake of completeness, let us sketch the derivation of the detector
pattern (\ref{eq: risposta totale Virgo}) in the TT gauge. We emphasize
that hereafter we closely follow the papers \cite{key-14,key-19}.

The TT gauge can be extended to SGWs in Scalar Tensor Gravity too
\cite{key-3,key-11,key-14,key-19}. In the TT gauge, for a purely
massless SGW propagating in the positive $z$ direction, with the
interferometer located at the origin of the coordinate system with
arms in the $\overrightarrow{x}$ and $\overrightarrow{y}$ directions,
the metric perturbation is given by \cite{key-3,key-11,key-14,key-19}

\begin{equation}
h_{\mu\nu}(t-z)=\Phi(t-z)e_{\mu\nu}^{(s)},\label{eq: perturbazione scalare}
\end{equation}

where $\Phi\ll1$, $e_{\mu\nu}^{(s)}\equiv diag(0,1,1,0)$, and the
line element is

\begin{equation}
ds^{2}=-dt^{2}+dz^{2}+[1+\Phi(t-z)][dx^{2}+dy^{2}].\label{eq: metrica puramente scalare}
\end{equation}

To compute the response function for an arbitrary propagating direction
of the SGW one recalls that the arms of the interferometer are in
the $\overrightarrow{u}$ and $\overrightarrow{v}$ directions, while
the $x,y,z$ frame of (\ref{eq: metrica puramente scalare}) is adapted
to the propagating SGW. Thus, the spatial rotations of the coordinate
system (\ref{eq: rotazione}) and (\ref{eq: rotazione 2}) are needed.
In this way the SGW is propagating from an arbitrary direction to
the interferometer (see figure 3). The beam splitter is also put in
the origin of the new coordinate system (i.e. $u_{b}=0$, $v_{b}=0$).
By using eq. (\ref{eq: rotazione}), eq. (\ref{eq: rotazione 2})
and eq. (\ref{eq: trasformazione metrica}), the line element (\ref{eq: metrica puramente scalare})
in the $\overrightarrow{u}$ direction becomes:

\begin{equation}
ds^{2}=-dt^{2}+[1+(1-\sin^{2}\theta\cos^{2}\phi)\Phi(t-u\sin\theta\cos\phi)]du^{2}.\label{eq: metrica scalare lungo u 2 bis}
\end{equation}

In this case, by applying the condition for null geodesics ($ds^{2}=0$)
in eq. (\ref{eq: metrica scalare lungo u 2 bis}) one gets immediately
eq. (\ref{eq: relazione spazio tempo}) because in the TT gauge the
coordinate time is exactly the proper time. In the same way, the line
element (\ref{eq: metrica puramente scalare}) in the $\overrightarrow{v}$
direction becomes:
\begin{equation}
ds^{2}=-dt^{2}+[1+(1-\sin^{2}\theta\sin^{2}\phi)\Phi(t-v\sin\theta\sin\phi)]dv^{2}.\label{eq: metrica + lungo v bis}
\end{equation}

At this point, one can perform in detail \textbf{exactly} the same
analysis in section 5 in order to obtain the detector pattern (\ref{eq: risposta totale Virgo}).

\end{document}